\documentclass[jap,twocolumn,reprint,showpacs,superscriptaddress,floatfix]{revtex4-1}
\usepackage{graphicx}
\usepackage{epsfig}
\usepackage{epstopdf}
\begin{document}

\title{Effect of composition on magnetocaloric properties of Mn$ _{3} $Ga$ _{(1-x)} $Sn$ _{x} $C}

\author{E. T. Dias},
\address{Department of Physics, Goa University, Goa 403206, India}
\author{K. R. Priolkar} \email[corresponding author:]{krp@unigoa.ac.in}
\address{Department of Physics, Goa University, Goa 403206, India}
\author{$\rm\ddot{O}$. $\rm\c{C}$akir}
\address{Physics Department, Yildiz Technical University, TR-34220 Esenler, Istanbul, Turkey}
\author{M. Acet}
\address{Faculty of Physics and CENIDE, Universitat Duisburg-Essen, D-47048 Duisburg, Germany}
\author{A. K. Nigam}
\address{Tata Institute of Fundamental Research, Dr. Homi Bhabha Road, Colaba, Mumbai 400005, India}
\date{\today}

\begin{abstract}
A study investigating the effect of Sn substitution on the magnetocaloric properties of Mn$ _{3} $Ga$ _{(1-x)} $Sn$ _{x} $C compounds reveals that nature of the MCE has a strong dependence on the nature of the magnetic ordering. For small amounts of Sn ($ x\leq0.2 $) the MCE is of the inverse type wherein an increase in the applied field beyond 5T gives rise to a table like temperature dependence of the entropy due to a coupling between the first order FM - AFM transition and the field induced AFM - FM transition. Replacement of Ga by larger concentrations of Sn ($ x\geq0.71 $) $(\Delta S_M)$ results in a change of the MCE  to a conventional type with very little variation in the position of $(\Delta S_M)_{max}$ with increasing magnetic field. This has been explained to be due to the local strain introduced by A site ions (Ga/Sn) which affect the magnetostructural coupling in these compounds. 
\end{abstract}

\maketitle

\section{Introduction}
The magnetocaloric effect $\left(MCE\right)$  intrinsic to all magnetic materials is due to the coupling of the magnetic sublattice with the magnetic field \citep{Pecharsky1999200} and can be characterized directly by measuring adiabatic change in temperature $\left( \Delta{T}_{ad} \right)$ upon application of magnetic field \citep{Bourgault201010, Amaral2010322} or  indirectly from heat capacity measurements or by estimating the isothermal entropy changes $\left( \Delta{S}_{M} \right) $ due to changes in applied magnetic field. Although this effect has been historically used in production of low temperatures, the discovery of giant magnetocaloric effect in Gd$_{5}$Si$_{2}$Ge$_{2}$ \citep{Pecharsky199778} has stimulated both basic and applied interest in the development of new materials that are useful for room temperature magnetic refrigeration as an alternative to vapor-compression technology \citep{Bruck200831}.

For the purpose of practical application, a magnetocaloric material must have comparatively large MCE values in small fields $ \left(H\leq 2T \right) $ accompanying a first order transition  preferably near room temperature along with a minimum thermal hysteresis  to ensure the life span of the magnetic refrigerator upon cyclically applying and removing external magnetic fields \citep{Pecharsky2009321}. Apart from Gd$_5$(Si$_x$Ge$_{1-x}$)$_4$, other known systems with considerable MCE values around room temperature include lanthanum manganese perovskite oxides \citep{Zhang199669}, transition metal based alloys \citep{Annaorazov199232, Tegus2002319}, Heusler alloys \citep{Hu200076, Hu200164, Planes200921} and 3d transition metal rich systems with values comparable to those of Gd\citep{Hu2001641, Hu200178}. These results including MnFeP$_{1-x}$As$_{x}$ system exhibiting reversible giant magnetic entropy change with the same magnitude as Gd$_{5}$Si$_{2}$Ge$_{2}$ \citep{Tegus2002415} renewed interest in materials exhibiting considerable magnetic entropy change at temperatures corresponding to their first order magnetic transitions \citep{Tegus2002415, Hu200280}.

Another class of materials that have attracted attention as candidates for ferroic cooling applications are the Mn based antiperovskite materials. Amongst these  Mn$_{3}$GaC undergoes a volume discontinuous first order transition  from a ferromagnetic (FM) to an antiferromagnetic (AFM) ground state at 160K  \cite{Bouchaud196637, Fruchart197844} accompanied by a large MCE in relatively low fields \cite{Garcia198315} as well as a table like MCE in high fields preferred when designing a practical refrigerant unit \cite{Tohei200394}.

On the other hand, Mn$_{3}$SnC exhibits a sharp first order change from a paramagnetic (PM) state to a non collinear ferrimagnetic (FIM) state at $\sim$ 279K accompanied by a conventional magnetic entropy change (80.69 mJ/cm$^{3}$K and 133 mJ/cm$^{3}$ K under a magnetic field of 2T and of 4.8T) \cite{Fruchart197844, Wang200985}. These materials are not only interesting from a technological point of view but also to understand the nature of magneto-structural coupling present in these antiperovskites. Prior to the first order transformation from a low volume cubic FM phase to high volume cubic AFM phase, Mn$_3$GaC undergoes a second order PM to FM transition at about 248K. The relative strengths of these two transitions can be modulated by controlling the carbon stoichiometry \cite{Dias2014363} or even by partial replacement of C by nitrogen \cite{Cakir2013344}. These studies indicate that the magneto-structural coupling is critically dependent on Mn $3d$ - C $2p$ band hybridization. Therefore a drastic change in nature of magnetic order via a replacement of Ga by Sn invokes interest. It has been recently shown that systematic replacement of Ga by Sn results in compounds showing characteristics of both the parent compounds \cite{Dias20141}. Another interesting aspect of Mn$_3$GaC is the existence of a completely reversible magnetic field induced first order transition from the high volume AFM state to a low volume FM state \cite{Cakir2014115}. Such a transition is absent in Mn$_3$SnC. As MCE is intimately related to the magneto-structural coupling, here we use it as a tool to understand the nature of magneto-structural coupling in Mn$_3$GaC and Mn$_3$SnC. Isothermal magnetic entropy changes under different applied magnetic fields, in the vicinity of the first order transitions have been studied in several compositions of Mn$_3$Ga$_{1-x}$Sn$_x$C ($0 \le x \le 1$). The results suggest that the magnetostructural coupling in these antiperovskite compounds critically depends on the local strain introduced by the A site cation.

\section{Experimental}
Isothermal magnetic entropy change associated with a magnetic transition of a material can be calculated from the field dependence of magnetization \cite{Gschneidner200030}. For this purpose Sn doped Mn$ _{3} $GaC type polycrystalline compounds of general formula Mn$ _{3} $Ga$ _{(1-x)} $Sn$ _{x} $C ($0 \leq x \leq 1$) prepared using the solid state reaction method and characterized for their structural, transport and magnetic properties as described in Ref.\cite{Dias20141} were used. In order to estimate isothermal magnetic entropy change $\Delta$S$ _{M} $ utilizing Maxwell's equation

\begin{equation}
\left(\frac{\partial S(T, H)}{\partial H} \right)_{T}  =\left( \frac{\partial M(T, H)}{\partial T}\right)_{H}
\label{eqn1}
\end{equation}

Which on integration for an isothermal-isobaric process gives,

\begin{equation}
\Delta{S}_{M}(T, \Delta{H})= \int\limits_{H_{1}}^{H_{2}}\left( \frac{\partial M(T, H)}{\partial T}\right)_{H} dH
\label{eqn2}
\end{equation}

 isothermal magnetization curves M(H) were recorded at several temperatures with an interval of 3K to 5K around the first order transition region \cite{Morrish1965}.
 
 The area between two magnetic isotherms recorded at close temperatures obtained from Eqn. \ref{eqn2} corresponds to a change in free energy which on dividing by the temperature interval gives the value of magnetic entropy change corresponding to an average of the two temperatures \cite{Amaral2010322}. Magnetization data for Mn$ _{3} $Ga$ _{(1-x)} $Sn$ _{x} $C ($0 \leq x \leq 1$) compounds was recorded using a SQUID magnetometer at temperature intervals of 3K to 5K. Each measurement as a function of field (0-7T) was carried out in zero field cooled (ZFC) mode wherein the sample was first heated to temperatures above its T$ _{C} $ in the PM state and then cooled to the measuring temperature in the absence of applied field.

\section{Results}
\begin{figure}
\begin{center}
\includegraphics[width=\columnwidth]{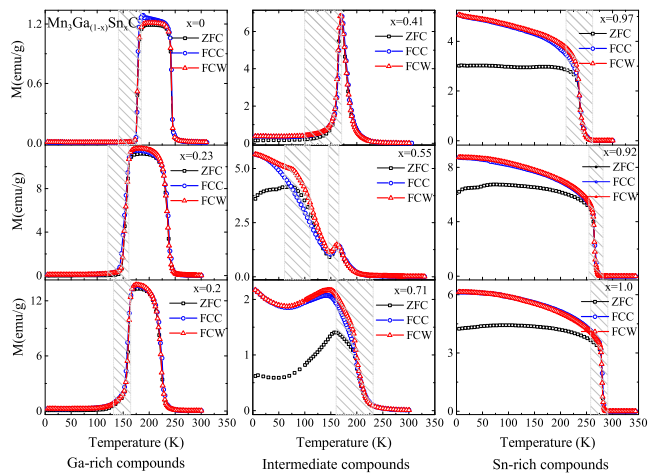}
\caption{Temperature dependence of magnetization measured in 0.01T applied field during ZFC, FCC and FCW cycles for Mn$_{3}$Ga$_{(1-x)}$Sn$_{x}$C.}
\label{fig:mt}
\end{center}
\end{figure}

A detailed analysis of structural and magnetic properties of Mn$ _{3} $Ga$ _{(1-x)} $Sn$ _{x} $C ($0 \leq x \leq 1$) compounds suggests that Sn doping in Mn$_3$GaC, though does not alter the primitive cubic crystal structure, leads to increase in unit cell volume along with a gradual increase in the strength of FM interactions to a point where the first order transition in Mn$_{3}$GaC is completely altered from FM$-$AFM type to a PM$-$FIM type in Mn$_{3}$SnC \cite{Dias20141}.  This behaviour can also be seen in Figure \ref{fig:mt}, wherein temperature dependence of magnetization measured in 0.01T applied field during zero field cooled (ZFC) and field cooled (FC) cycles is plotted for several compositions of Mn$_{3}$Ga$_{(1-x)}$Sn$_{x}$C. With decreasing temperature the Ga rich compounds, $x \leq 0.23$, first transform from a room temperature PM state to an intermediate FM state via a second order transition before undergoing a first order transition to the AFM ground state. M(T) results for compounds with intermediate concentrations of Ga and Sn (0.4 $\leq$ x$\leq$0.7) show a co-existence of  two magnetic phases: Ga rich phase which orders antiferromagnetically and a Sn rich phase with dominant ferromagnetic interactions. The strength of FM interactions continues to further  increase across the series while suppressing the AFM state in the Sn rich ($ x \geq 0.8$) compounds.

\begin{center}
\begin{figure}
\includegraphics[width=\columnwidth]{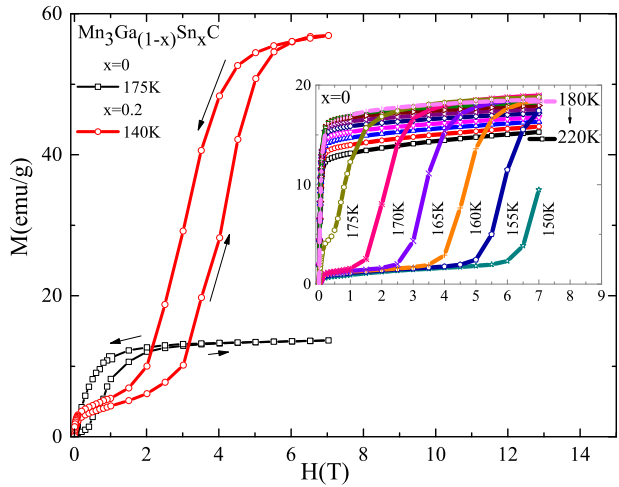}
\caption{Isothermal magnetization curves as a function of external magnetic field, H $\le$ 7T  for Mn$ _{3} $Ga$ _{(1-x)} $Sn$ _{x} $C, ($x = 0$ and $x = 0.2$) compounds  measured across the first order transition region.}
\label{fig:mh1}
\end{figure}
\end{center}

\begin{center}
\begin{figure}
\includegraphics[width=\columnwidth]{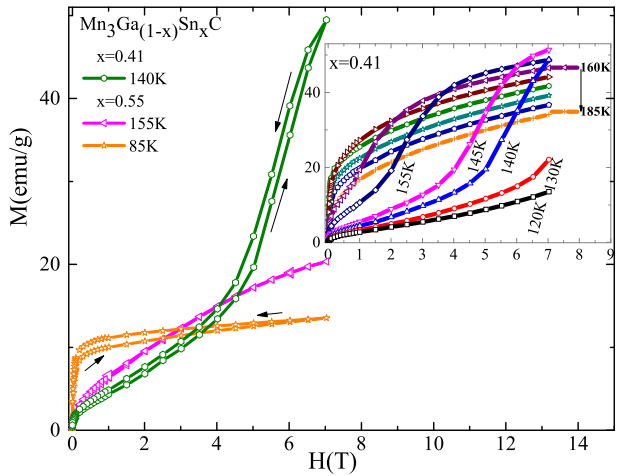}
\caption{Magnetization curves M(H), measured across the first order transition of Mn$ _{3} $Ga$ _{(1-x)} $Sn$ _{x} $C with $x$ = 0.41 and 0.55 as a function of external magnetic field (H $\le$ 7T).}
\label{fig:mh2}
\end{figure}
\end{center}

\begin{center}
\begin{figure}
\includegraphics[width=\columnwidth]{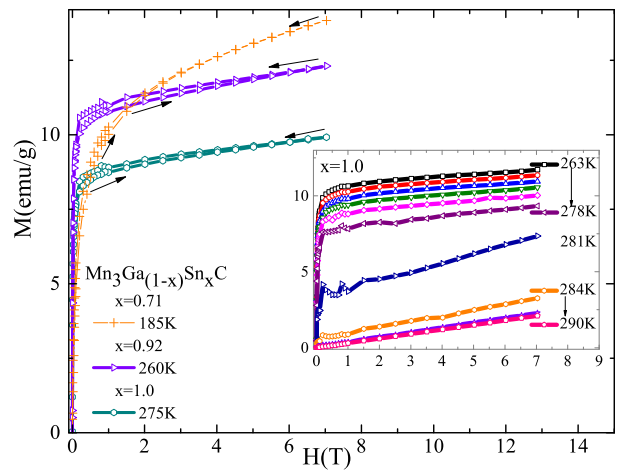}
\caption{M(H) curves as a function of external magnetic fields (0 $\le$ H $\le$ 7T)  for Sn-rich Mn$ _{3} $Ga$ _{(1-x)} $Sn$ _{x} $C compounds ($x \ge 0.71 $) measured across the first order transition region.}
\label{fig:mh3}
\end{figure}
\end{center}

The change in the type of magnetic interactions present in magnetic ground state of Mn$_3$Ga$_{1-x}$Sn$_x$C compounds is further illustrated by magnetization isotherms M(H) recorded in the vicinity of their respective first order transitions and presented in Figures \ref{fig:mh1}, \ref{fig:mh2} and \ref{fig:mh3}. As can be seen in Fig. \ref{fig:mh1}, M(H) for Mn$_3$GaC ($x = 0$) recorded at 175K exhibits a typical ferromagnetic like behavior even though the compound is in AFM state ($T_N$ = 179K). The reason for such a behavior is clear from the magnetization curves present in the inset of Fig. \ref{fig:mh1}. Mn$_3$GaC undergoes a field induced metamagnetic transition from antiferromagnetic to ferromagnetic state. The critical field value of this transition of course increases with lowering of temperature. It may be pointed out that neutron diffraction studies have shown that not only the magnetic order but also the unit cell volume returns to its pre-transition value \cite{Cakir2014115}. This is a clear indication of coupling between magnetic and structural degrees of freedom.

For Sn doped compounds ($x \leq 0.55$), though the metamagnetic transition still occurs, the magnetic field required to induce this transition increases with Sn concentration. For example in $x = 0.2$, the transition occurs at H $\sim$ 3T at 140K (T/T$_N$ $\sim$ 0.95) while in $x = 0$ only 1T magnetic field is required to induce the transition at the same relative temperature (T = 170K). Furthermore, this value of magnetic field increases to $\sim$ 5T in $x = 0.41$ and perhaps to more than 7T in $x = 0.55$. For higher Sn concentration no metamagnetic transition is seen and the magnetic order is dominated by ferromagnetic interactions. In fact even in case of $x = 0.55$ wherein there are two first order magnetic transitions at 155K and 85K and which have been respectively attributed to Ga rich and Sn rich regions \cite{Dias20141}, the transition at 85K does not exhibit any signatures of a metamagnetic transition. At higher concentrations of Sn, FM interactions appear to strengthen and the transition temperature increases to $\sim$ 280K in Mn$_3$SnC. Such a behavior indicates that Sn doping not only strengthens the ferromagnetic interactions but at the same time weakens the AFM state. Local interactions usually play a major role in metamagnetic transitions occurring in the vicinity of a first order transition \cite{Takenaka201415}. Therefore it is quite intriguing to understand how Sn doping, beyond a critical concentration, changes the magnetic character of the compounds, increases the transformation temperature while completely destroying the AFM state.

\begin{figure}
\begin{center}
\includegraphics[width=\columnwidth]{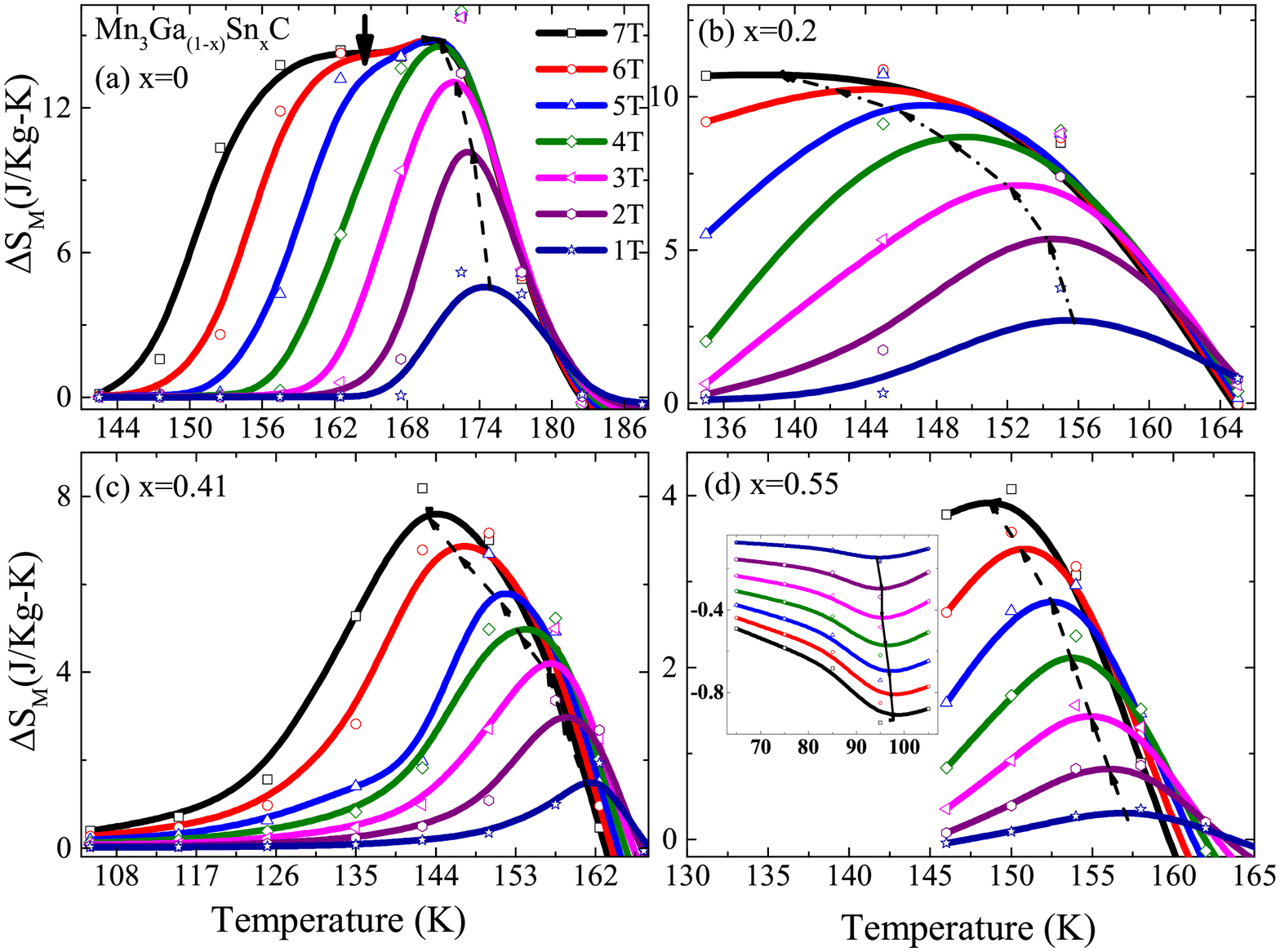}
\caption{Entropy change in the vicinity of the magnetostructural transition as determined from magnetization measurements for Mn$ _{3} $Ga$ _{(1-x)} $Sn$ _{x} $C, $ 0 \le x \le 0.55$}
\label{fig:mce1}
\end{center}
\end{figure}

\begin{figure}
\begin{center}
\includegraphics[width=\columnwidth]{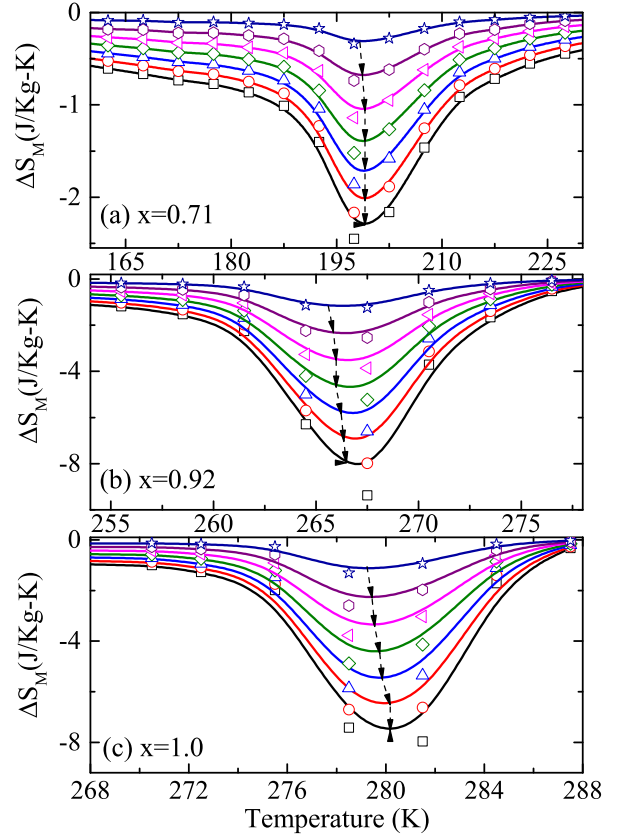}
\caption{Entropy change in the vicinity of the magnetostructural transition as determined from magnetization measurements for Sn rich Mn$ _{3} $Ga$ _{(1-x)} $Sn$ _{x} $C }
\label{fig:mce2}
\end{center}
\end{figure}

We seek answers to the above through a study of magnetic entropy behaviour in Mn$_{3}$Ga$_{(1-x)}$Sn$_{x}$C. Figures \ref{fig:mce1} and \ref{fig:mce2} present a graphical representation of temperature dependence of entropy change associated with first order phase transition for Mn$_{3}$Ga$_{(1-x)}$Sn$_{x}$C (0$ \leq $x$ \leq $1) compounds. $\Delta S_{M}$ values were calculated from magnetization isotherms near the ordering temperature using equation \ref{eqn2}. As expected, the rapid change in magnetization seen at the first order transition in these materials is accompanied by maximum entropy change indicated by prominent peak like structures around the transition regions marked in Figure \ref{fig:mt}. Though the transition is clearly of first order for all compounds belonging to the series, the nature of the MCE has a strong dependence on the Sn content.

The entropy change in Ga rich compounds accompanying the first order FM-AFM transition is depicted by inverse MCE peaks in Figures \ref{fig:mce1}a and \ref{fig:mce1}b. In Mn$_3$GaC ($x$ = 0), as the magnetic field increases, a second peak (indicated by an arrow in Figure \ref{fig:mce1}a) develops (for H $\ge$ 5T). Such a two peak structure has been reported in compounds with a coupled first order magnetic transition and a metamagnetic transition \cite{Liu2014105}. The observed profile in Figure \ref{fig:mce1}a could also indicate a simple broadening of the peak for H $\ge$ 5T indicating that the whole transition is induced for this range of magnetic fields. However, a clear shift in the position of the peak towards lower temperatures with increasing magnetic field can be seen in the figure.  The calculated maximum values of $\Delta S_{M}$ at the AFM ordering temperature and the associated adiabatic temperature difference of about $\Delta$T$  \sim$4K are in fair agreement with  reported values \cite{Dias2014363}. A similar nature of $\Delta S_M$ versus temperature profile can be seen in sample with 20\% Sn doping. Here too, the applied field shifts the first order transition towards lower temperatures but the $\Delta S_{M}$ value, in comparison to undoped compound is lower at 7.4 J/kg-K (H = 2T) and attains a maximum value of 10.8 J/kg-K in an applied field of 7T. Increasing the magnetic field beyond 5T does not seem to affect the maximum MCE value, but gives rise to a table like temperature dependence of the entropy due to a coupling between the first order FM - AFM transition and  the field induced AFM - FM transition \citep{Lewis200393}.

Entropy changes as a function of temperature in various applied fields shown in Figures \ref{fig:mce1}c, \ref{fig:mce1}d and Figure \ref{fig:mce2}a  for three intermediate concentrations of Sn depict a strong dependence of the magnitude of $\Delta S_{M}$ on the nature of magnetic ordering in a material. In the case of $x$ = 0.41 also the maximum of $\Delta S_M$, ($(\Delta S_M)_{max}$), shifts to lower temperatures with increasing magnetic fields. Furthermore, the increasing width of the peak also hints towards a coupling seen in Ga rich compounds. A similar behavior is also noticed for the 155K transition in $x$ = 0.55 (Figure \ref{fig:mce1}d). However, the first order transition at 85K in $x$ = 0.55 neither exhibits broadening nor a shift in the $(\Delta S_M)_{max}$. This transition at 85K is ascribed to the ordering of Sn rich regions, while the transition at 155K is related to transformation of Ga rich regions \cite{Dias20141}. The fact that $(\Delta S_M)_{max}$ at 155K shows a variation in its position as a function of magnetic field and the entropy peak at 85K does not, indicates the changing nature of magnetostructural coupling due to Sn doping. This is further confirmed from the behavior of $\Delta S_M$ versus temperature curves calculated for different magnetic fields in $x$ = 0.71, 0.92 and 1.0 which are presented in Figures \ref{fig:mce2}a, \ref{fig:mce2}b and \ref{fig:mce2}c. In all these cases not only the $\Delta S_M$ is negative indicating conventional magnetocaloric effect, but also there is very little variation in the position of $(\Delta S_M)_{max}$ with increasing magnetic field.

\section{Discussion}
Magnetization measurements reveal that doping of Sn in Mn$_3$GaC results in evolution of FM interactions in the transformed AFM phase of Ga rich compounds along with lowering of the first order transformation temperature. The $x$ = 0.55 compound undergoes two first order transformations respectively at 155K and 85K which are ascribed to Ga rich regions and Sn rich regions. With further increase in Sn concentration, the first order transition temperature rises steeply and the compounds transform from a PM to FIM state. Another notable feature is the disappearance of the field induced transition with Sn doping. Field induced AFM to FM transition is present in all Ga rich compounds ($x \le 0.41$) but the critical field required to induce this transition increases with Sn concentration.

The metamagnetic transition in Mn$_3$GaC occurs from a high volume AFM state to low volume FM state\citep{Fruchart197844}. The replacement of Ga by larger Sn results in a natural volume expansion which not only lowers the transformation temperature but also pushes the metamagnetic transition to higher magnetic fields. These two magnetostructural transitions in Mn$_3$GaC are magneto-volume transitions wherein the magnetic field plays a role akin to that of hydrostatic pressure. This is clear from the fact that in all Ga rich, Mn$_3$Ga$_{1-x}$Sn$_x$C ($x \le 0.41$) compounds the first order low volume FM to high volume AFM transition shifts to a lower temperature in presence of magnetic field. The increase in the critical field required to induce AFM-FM metamagnetic transition with increasing Sn content could therefore be related to the increase in unit cell volume with Sn doping. The larger the volume of the unit cells, higher the magnetic field (more pressure) required to convert them from the high volume AFM to low volume FM state. 
 
\begin{figure}
\begin{center}
\includegraphics[width=\columnwidth]{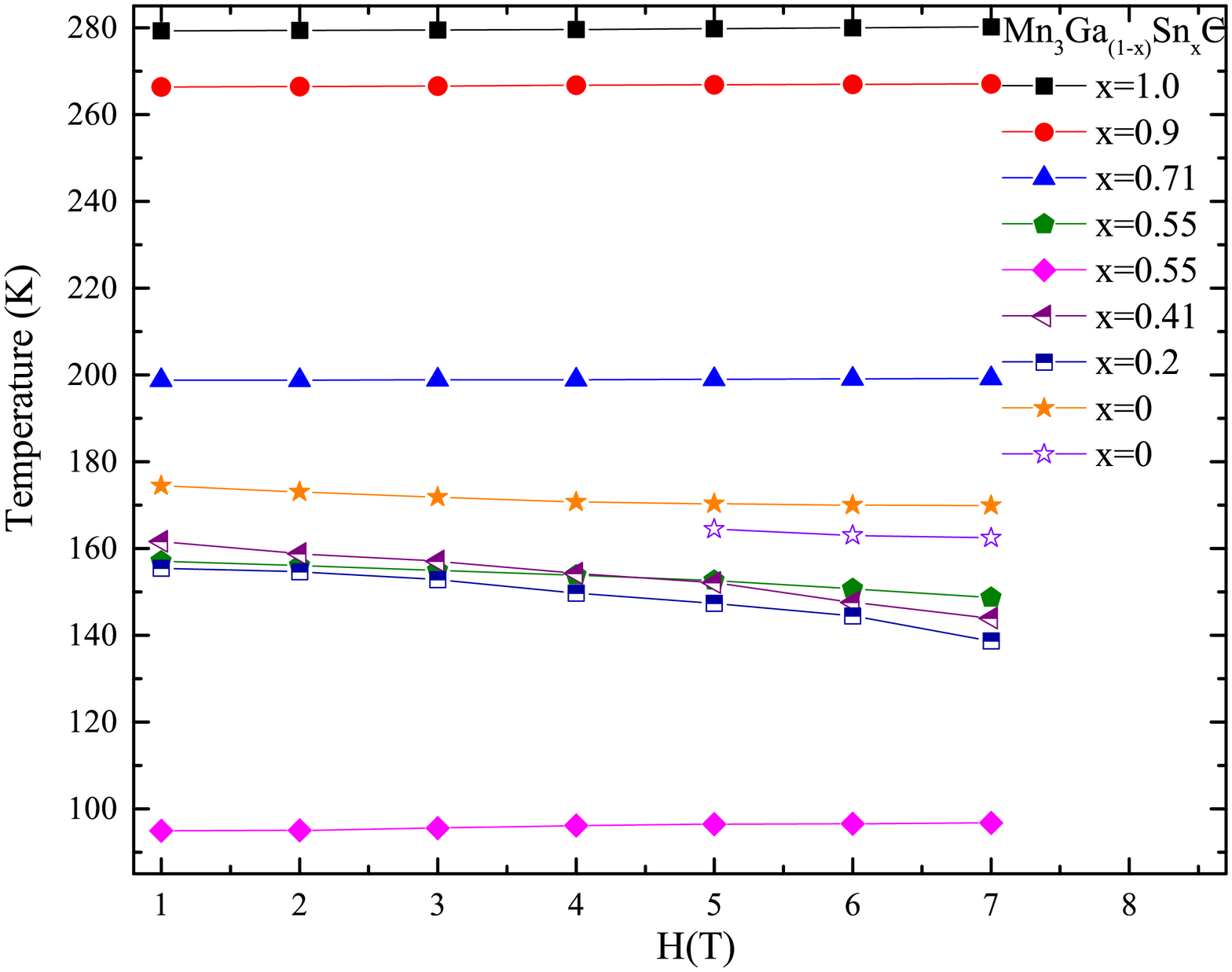}
\caption{Variation of $\Delta T$ as a function of applied magnetic field for Mn$ _{3} $Ga$ _{(1-x)} $Sn$ _{x} $C compounds.}
\label{fig:mceh}
\end{center}
\end{figure}

In $x$ = 0.55, two first order transitions at 155K and 85K, corresponding respectively to Ga rich regions and Sn rich regions are observed. The transition at 155K shows a similar magnetic field dependence as other Ga rich compounds even though it occurs at a slightly higher temperature compared to that in $x$ = 0.41. While the transition at 85K exhibits negligible field dependence thereby supporting the argument of existence of two magnetic phases \cite{Dias20141}. This behaviour can be clearly seen from the variation in temperature of peak of $(\Delta S_M)$ ($T_p$) as a function of magnetic field. In Figure \ref{fig:mceh}, the plot of $\Delta T = T_p(H) - T_P(0.1T)$ as a function of magnetic field shows a large variation for Ga rich compounds. In the case of $x$ = 0.55, the transition at 155K exhibits variation in its position with increasing magnetic field while the transition at 85K has a behaviour similar to that exhibited by Sn rich compounds. 

Up to about $x < 0.5$, the role of Sn could be that of increasing the lattice volume. For higher Sn concentrations, its role appears to change. Inspite of a steady increase in the lattice constant from Mn$_3$GaC to Mn$_3$SnC, the first order transformation temperature increases with increasing Sn concentration for $x \geq 0.5$. This is exactly opposite to the behavior of Ga rich Mn$_3$Ga$_{1-x}$Sn$_x$C wherein the substitution of Sn causes the transformation temperature to decrease due to cell volume expansion. Alternately, starting with Mn$_3$SnC the decrease in first order transformation temperature due to Ga substitution can be ascribed to a decrease in unit cell volume. In itinerant electron magnetic systems, a decrease in cell volume increases the band width which suppresses magnetic moment \cite{Wasserman19905, Yoshinori200618}. Therefore as the Sn content is diluted with smaller Ga, the magnetic transformation temperature decreases. Again this explanation is valid only up to 50\% Ga doping. For higher Ga concentrations, there is not only an abrupt increase in transformation temperature from 85K to $\sim$ 150K, the nature of transformation also changes from PM$-$FIM type to FM$-$AFM type. This behavior points to the changing role of the A site ion in the magneto-structural transformation in antiperovskites. Such a crucial role of A site ions has also been highlighted in another family of Mn antiperovskites containing nitrogen \cite{Takenaka201415}.

Though both, Mn$_3$GaC and Mn$_3$SnC are isostructural compounds undergoing a volume discontinuous first order transformation, there are many dissimilarities between them. Apart from the difference in nature of magnetic transition associated with the structural transition in the two compounds, the magnetic propagation vector as indicated by neutron diffraction studies is also different \citep{Fruchart197844} . In Mn$_3$GaC, the AFM propagation vector is along the [$ \frac{1}{2} $, $ \frac{1}{2} $, $ \frac{1}{2} $] direction giving rise to a collinear AFM order along the [111] direction \cite{Fruchart19708}. While in Mn$_3$SnC, the AFM order propagation vector, $\vec k$ = [$ \frac{1}{2} $, $ \frac{1}{2} $, $ 0 $] resulting in an anisotropic structure of the type (a$ \sqrt{2} $, a$ \sqrt{2} $, a) with a square arrangement of the spins at two of the Mn sites giving rise to an  AFM  configuration with a small ferromagnetic component and a parallel FM component at the third site responsible for the FM like behavior \cite{Heritier197712}.

Recent EXAFS studies have shown that the CMn$_6$ octahedra in Mn$_3$SnC is locally distorted even in the paramagnetic phase and inspite of a cubic crystal structure \cite{ed}. The distortion of CMn$_6$ octahedra gives rise to shorter and longer Mn-Mn bond distances which critically affect the competition between nearest Mn-Mn antiferromagnetic interactions and next nearest Mn-Mn ferromagnetic interactions \citep{Takenaka201415} and has been considered to be the reason for the change in the direction of magnetic propagation vector in Mn$_3$SnC with respect to that in Mn$_3$GaC.
Such a distortion can also be used to explain the observed behavior of magnetocaloric effect and change in the first order magnetic transformation temperature. Mn$_3$GaC consists of regular CMn$_6$ octahedra. Addition of Sn in place of Ga in Mn$_3$GaC results in a anisotropic tensile strain in some of the CMn$_6$ octahedra and therefore makes it that much harder for the applied magnetic field to induce a magneto-volume transformation which converts the AFM ground state to FM one. With increase in Sn concentration therefore, the metamagnetic transition shifts to higher values of applied magnetic field. In $x = 0.55$, the two types of octahedra, unstrained (Ga rich regions) and strained (Sn rich regions) could be nearly equal in number and hence two first order transformations are seen. With further increase in Sn content, the number of strained octahedra exceeds the number of unstrained ones and the temperature of first order  transformation increases. Therefore, the local strain introduced by the A site cation seems to play an important role in magneto-volume transformation in Mn$_3$Ga$_{1-x}$Sn$_x$C antiperovskite compounds.

\section{Conclusions}
In summary, the paper reports a study on the nature of magnetocaloric effect in Mn$_3$Ga$_{1-x}$Sn$_x$C antiperovskite compounds. It is seen that compounds rich in Ga content exhibit a strong magnetic field dependence of the FM-AFM transformation as well as a field induced magneto-volume transformation from AFM to FM state. Such a field dependence of first order transition as well as metamagnetic transition is absent in Sn rich compounds. This has been shown to be due to a local strain introduced by substitution of larger Sn in place of Ga. Presence of such a local strain introduces FM interactions through a change in magnetic propagation vector and hence the first order transformation temperature decreases. A suppression of metamagnetic transition is also seen with increasing Sn concentration. With increase in Sn concentration beyond $x = 0.5$, the higher number of strained CMn$_6$ octahedra causes the first order transformation temperature to increase with increase in Sn content.

\section{Acknowledgments}
Authors thank Board of Research in Nuclear Sciences (BRNS) for the financial support under the project 2011/37P/06. M/s Devendra D. Buddhikot and Ganesh Jangam are acknowledged for the experimental assistance.

\bibliographystyle{apsrev4-1}
\bibliography{references}

\end{document}